\documentclass[conference]{IEEEtran}

\ifCLASSOPTIONcompsoc
  \usepackage[nocompress]{cite}
\else
  \usepackage{cite}
\fi

\usepackage{cite}
\usepackage{amsmath}
\usepackage{amssymb}
\usepackage{amsfonts}
\usepackage{graphicx}
\usepackage{textcomp}
\usepackage{xcolor}
\usepackage{multirow}
\usepackage{soul}
\usepackage{tikz}
\usepackage{algorithm}
\usepackage{algpseudocode}
\usepackage{colortbl}
\usepackage{bm}
\usepackage{float}
\usepackage{filecontents}

\newcommand{\cmmnt}[1]{}  

\def\BibTeX{{\rm B\kern-.05em{\sc i\kern-.025em b}\kern-.08em
    T\kern-.1667em\lower.7ex\hbox{E}\kern-.125emX}}

\usepackage{todonotes}

\ifCLASSINFOpdf
\else
\fi

\begin{document}
\newcommand{\algrule}[1][.2pt]{\par\vskip.5\baselineskip\hrule height #1\par\vskip.5\baselineskip}

\title{\huge NeSe: Near-Sensor Event-Driven Scheme for Low Power Energy Harvesting Sensors\vspace{-0.5em}}


\author{Sepehr Tabrizchi$^\ast$, Mehrdad Morsali$^\dagger$, Shaahin Angizi$^\dagger$, Arman Roohi$^\ast$\\
\small
$^\ast$School of Computing, University of Nebraska–Lincoln, Lincoln NE, USA\\
$^\dagger$Department of Electrical and Computer Engineering, New Jersey Institute of Technology, Newark, NJ, USA\\
shaahin.angizi@njit.edu, aroohi@unl.edu}
\vspace{-1.5em}

\maketitle
\begin{abstract}
Digital technologies have made it possible to deploy visual sensor nodes capable of detecting motion events in the coverage area cost-effectively. However, background subtraction, as a widely used approach, remains an intractable task due to its inability to achieve competitive accuracy and reduced computation cost simultaneously. In this paper, an effective background subtraction approach, namely NeSe, for tiny energy-harvested sensors is proposed leveraging non-volatile memory (NVM). Using the developed software/hardware method, the accuracy and efficiency of event detection can be adjusted at runtime by changing the precision depending on the application's needs. Due to the near-sensor implementation of background subtraction and NVM usage, the proposed design reduces the data movement overhead while ensuring intermittent resiliency. The background is stored for a specific time interval within NVMs and compared with the next frame. If the power is cut, the background remains unchanged and is updated after the interval passes. Once the moving object is detected, the device switches to the high-powered sensor mode to capture the image. 

\end{abstract}


\section{Introduction}
From energy-harvested surveillance and monitoring systems in smart cities to smart human-machine interfaces in mobile devices, smart, low-power, connected sensors are attracting increasing interest in a wide variety of applications. Moreover, our environment can be best described through vision, which is becoming increasingly ubiquitous in video monitoring applications.
Human observers monitor several cameras to detect unusual activity and provide immediate feedback. Unfortunately, human observers lose 90\% of their concentration capability after only 20 minutes of following ten cameras attentively, which defeats the purpose of this approach.
Therefore, the automatic detection of unusual events in embedded applications is becoming increasingly significant. 
Machine vision applications often begin with background subtraction, making it an essential component. Inputs from background subtraction are given to higher-level processes, such as object tracking.
An online video background subtraction usually consists of two stages: initialization of the background model, in which the bootstrapping is performed, and the background model's maintenance, which involves updating the parameters online.
Interpreting a scene, however, requires large amounts of computing power and data-intensive vision algorithms.
As they are highly parallelizable, pixel-level foreground detectors are ideal for embedded platforms. 
CMOS imagers with on-chip feature extraction and compression have been developed extensively in the last decade with the primary goal of optimizing computing resources and reducing overall power consumption \cite{oliveira2013cmos,cevik2015ultra,cottini201333}.

In this work, we propose a near-sensor event-driven architecture, namely \emph{NeSe}, allowing for a trade-off between accuracy and power efficiency. NeSe is capable of operating in different modes, 12 in total, regarding the precision and box sizes, which will be explained in the following. To the best of our knowledge, this work is the first that utilizes non-volatile elements to store the static background, which leads to a notable reduction in standby power consumption.

\begin{figure*}[t]
\centering
\includegraphics [width=0.67\linewidth]{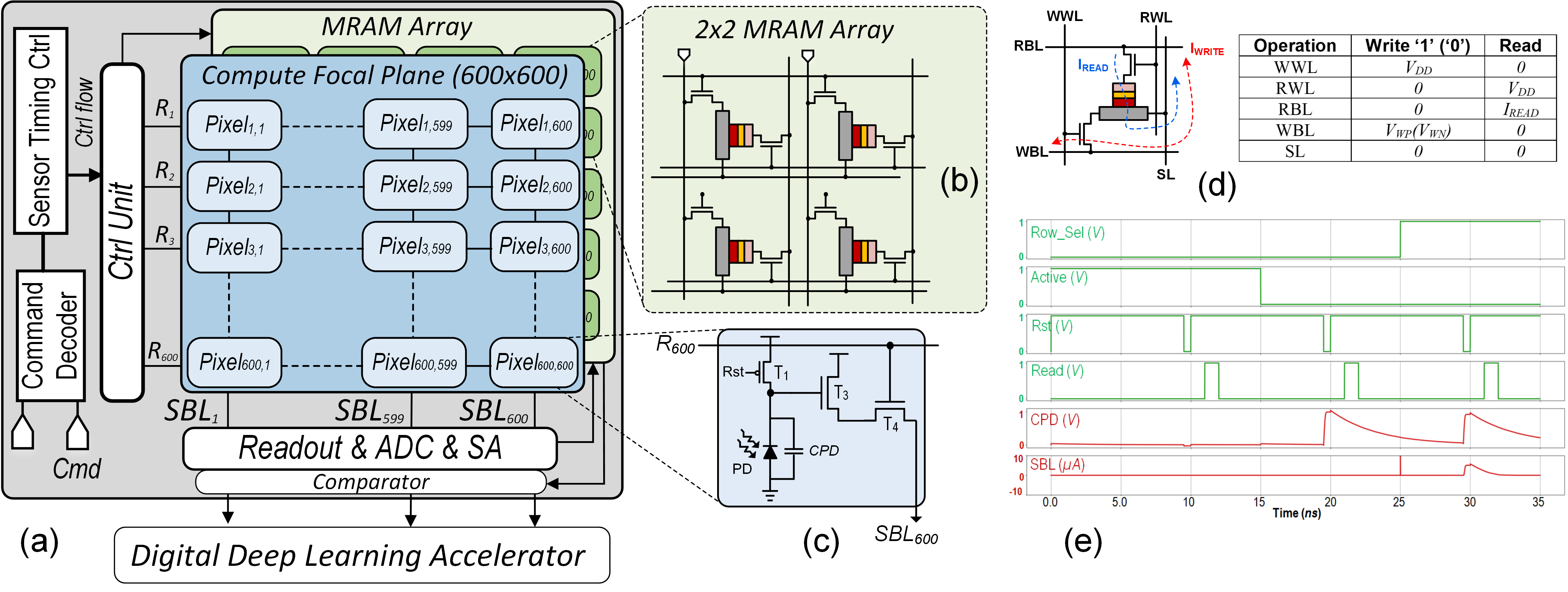} 
\vspace{-0.8em}
\caption{(a) The NeSe architecture, including (b) an MRAM array and (c) a pixel. (d) Schematic and biasing of an MRAM, and (e) pixel's transient waveform.}
\vspace{-1.75em}
\label{main}
\end{figure*}

\section{Near-Sensor Processing Background}
In the same way that eyes and brains work together, the sensors that detect the field of view generate a stream of pixels that represent the scenic event and are sent to a backend processor.
Although there are 130 million pixels on the retina, the brain only has 1.3 million synaptic connections, which indicates a high sparsity ratio. This massive sparsity can significantly reduce power consumption and latency. A further improvement can be made by reducing the amount of redundant information sent to the brain. 
By inspiring from the observations and taking steps to mitigate the abovementioned issues, the integration of computing and sensing has been extensively studied, reducing data movement and ADC bandwidth. 
The research outcomes are classified into three designs, 
processing-near-sensor (PNS) \cite{li2021ns,hsu20200}, processing-in-sensor (PIS) \cite{xu2020utilizing,angizi2022pisa,abedin2022mr}, and finally processing-in-pixel (PIP) \cite{xu2020macsen,tabrizchi2022ocelli}.
Most computer vision systems perform background subtraction as a first step in detecting moving objects within a video stream without having prior knowledge of the objects themselves \cite{garcia2020background}. 
A background model is generally created during the background subtraction process. The easiest way to do this is to manually set a static image for the background that has no moving objects. Each video frame is then compared to the static image to compute the absolute difference, referred to as Static Frame Difference, and is represented by: $\scriptstyle |F_i - B|> TH$.
In the event of changing ambient lighting, a static image may not be the best choice since the foreground segmentation may fail completely. 
Alternately, the previous frame may be used instead of a static image, referred to as Frame Difference, and is expressed by: $\scriptstyle |F_i - F_{i-1}|> TH$.
Due to its sensitivity to threshold TH, this technique may only work properly under certain frame rates and object speeds, e.g., it fails if the moving object stops suddenly. 
Thus, Authors in \cite{lai1998fast} modeled the background more accurately using the average, arithmetic mean, or weighted mean of several previous frames. The equation for the past $n$ frame is: $B_j= \frac{1}{n} \Sigma_{i=0}^{n-1} F_{j-i} $.
In order to store more frames in off-chip memory, this model requires high memory storage. Consequently, additional computations and memory accesses are needed, which conflict with resource-constrained tiny devices.
\vspace{-0.5em}

\section{Proposed Design}

We propose NeSe as an efficient and reconfigurable always-on intelligent visual perception architecture as shown in Fig. \ref{main}(a) that realizes a near-sensory processing scheme with event detection capabilities. 
NeSe consists of a $600\times 600$ pixel array (PA), a non-volatile Spin-Orbit Torque Magnetic Random Access Memory (SOT-MRAM) array, a row controller (Ctrl), a command decoder, a sensor timing Ctrl, a memory/computing unit, and readout/ADC/SA/comparator circuitry. The storage element in SOT-MRAM is SHE-MTJ \cite{fong2015spin}
Each cell located in the MRAM array is connected with a Write Word Line (WWL), Write Bit Line (WBL), Read Word Line (RWL), Read Bit Line (RBL), and Source Line (SL). The bit-cell structure of 2T1R SOT-MRAM and its biasing conditions are shown in Fig. \ref{main}(d).
The proposed architecture operates in two modes, i.e., \emph{sensing} and \emph{event detection}. 
The NeSe architecture captures the input as a static background and then writes the central pixels according to the configured settings into the MRAM cells. Due to the non-volatility feature of MRAMs, if the power of NeSe is cut, the initial background is still held. Once the moving object is detected, the architecture switches to the sensing mode to detect the object(s). To reduce the overall power consumption, NeSe only updates (sends) the modified pixels on the MRAM array.

\subsection{Pixel and MRAM arrays}
Illustrated in Fig. \ref{main}(c), conventional NeSe's pixel consists of a three-transistor/one-photodiode (PD) sensor. In the sensing mode, by initially setting Rst=`high', the PD connected to the T\textsubscript{1} transistor turns into inverse polarization and the readout component captures a $V_{1}$ = VDD voltage.
By turning off T\textsubscript{1}, PD generates a photo-current with respect to the external light intensity which in turn leads to a voltage drop ($V_{PD}$) at the gate of T\textsubscript{2}. Therefore, the voltage values before and after the image light exposure, i.e., $V_{1}$ and $V_{2}$, are sampled by the readout circuit, and the difference between the two voltages is sensed, amplified, and then converted to digital data by an ADC. This value is proportional to the voltage drop on $V_{PD}$. Figure \ref{main}(e) depicts the functionality of one proposed pixel.
It is worth pointing out that each ADC samples when the voltage drops, then it subtracts the pixel reset voltage and converts the output signal. Accordingly, the ADC can skip to the next row of the array. 
NeSe is equipped by a near-sensor CMOS bit-wise XNOR comparator, as shown in Fig. \ref{main}(a), to efficiently compare such row-wise digitized pixel data with the corresponding captured background in the MRAM array to detect events. To enable this, one row of the MRAM array, shown in Fig. \ref{main}(b), is selected, sensed out, and loaded as the first operand into a register at the comparator where the second register is loaded by the pixel data. Accordingly, a single-cycle XNOR operation is accomplished. If a mismatch is detected, i.e., an event observed, the MRAM array holding the central pixels requires to be updated. Computationally, this stage requires \textit{n} 
MRAM write operation. To achieve an ultra-fast low-energy write operation, the SOT-MRAM cells are developed with a 20$K_bT$ energy barrier. As experimentally shown in \cite{Roohi2018TC}, this will reduce the write energy consumption by half compared with the conventional 40$K_bT$ design at the cost of lower retention time.   
\vspace{-0.5em}

\subsection{Event-Detection Mode}
The primary task of the always-on NeSe architecture is to detect an event using background variations.
NeSe supports 12 various implementations to consider both efficiency and accuracy design metrics. Different designs are determined by the $box\_size \in \{3,5,7\}$ and $precision \in \{1,2,3,4\}$, where $box\_size$ represents height and width of defined groups, and $precision$ denotes the bit-width of ADCs. Each $n\times n$ pixel box includes only one ON pixel, ($n-1$) Disconnect pixels, and ($n^2-n$) OFF pixels. An implementation with a larger box size reduces power consumption at the cost of accuracy degradation.
In NeSe, each column is enabled via a distinct but common $V_{DD}$, and each row is chosen using a common row selector (\emph{R}) signal. Thus, the ON, Disconnect, and OFF pixels are formed when \emph{R} and the column are enabled, \emph{R} is disabled, but the column is enabled, and the column is disabled regardless of the \emph{R} value, respectively. The \emph{R} signal is valued using ($nx - 1$), where $n \in \{3,5,7\}$ and $x$ is the row index $\in \{1,2,\dots, \lfloor 600/n \rfloor \}$.  
\begin{figure}[t]
\centering
\includegraphics [width=0.95\linewidth, height=4.75cm]{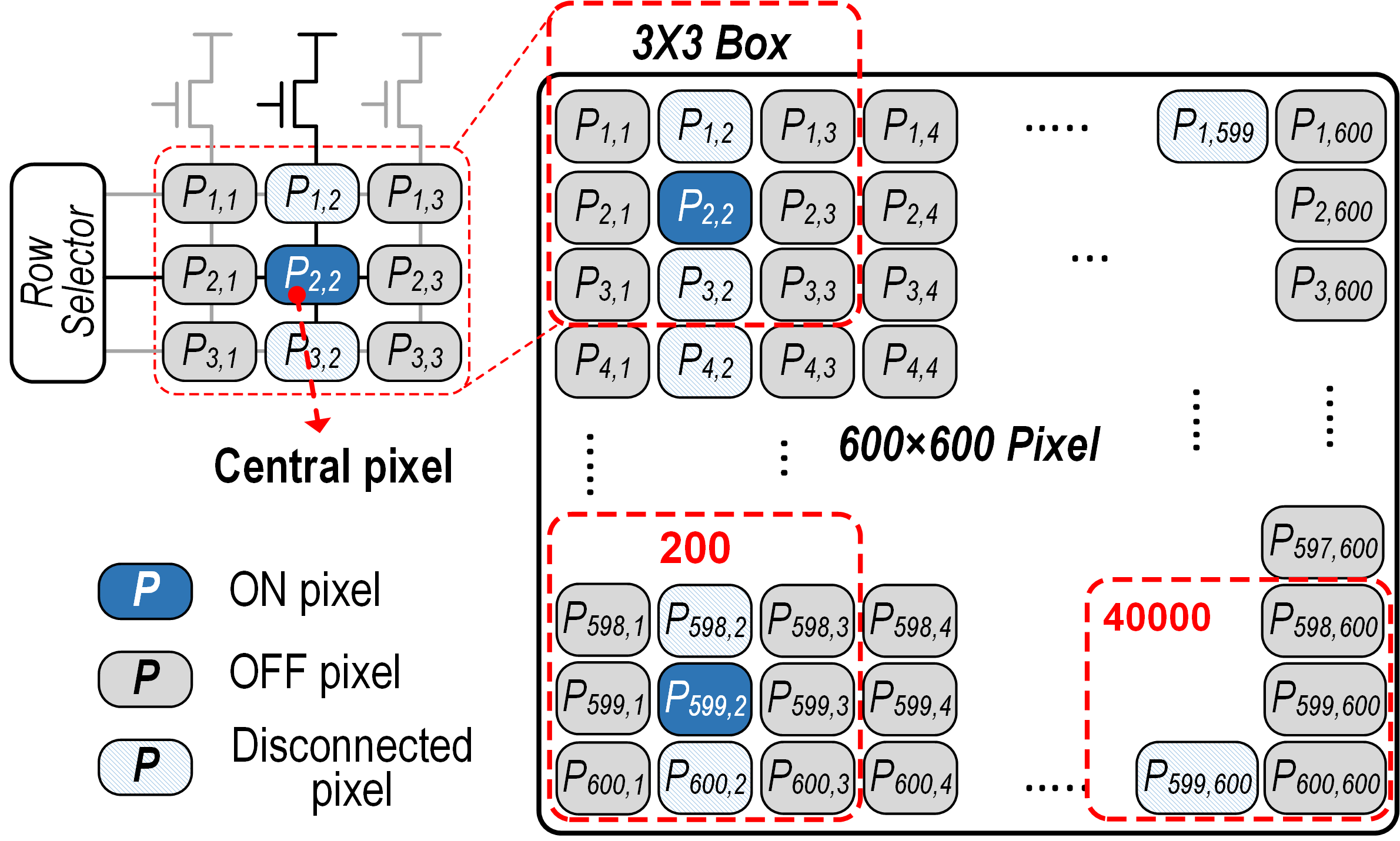} 
\vspace{-0.75em}
\caption{Boxing pixels with the size of $3 \times 3$ with the possible situations.}
\label{pixel}
\end{figure}
Consequently, all the columns without central pixels are disconnected from the power supply (OFF), while the rest of the pixels in the columns containing the central pixel is disconnected using \emph{R} signal.
For instance, as shown in Fig. \ref{pixel}, by setting $box\_size$ to 3, all the pixels are grouped into a $3\times 3$ shape, where the central pixel (e.g., $P_{2,2}$) is ON, other two pixels (e.g., $P_{1,2}$ and $P_{3,2}$) in the same column are disconnected from ADCs because of \emph{R} values, and the rest (e.g., $P_{1,1}, P_{2,1}, P_{3,1}, P_{1,3}, P_{2,3}$ and $P_{3,3}$) are OFF.

\begin{table}[b]
\centering
\caption{Effect of box size in NeSe properties.}
\scriptsize
\begin{tabular}{cccccc}
\hline
\multirow{2}{*}{\begin{tabular}[c]{@{}c@{}}\textbf{Box}\\ \textbf{Size}\end{tabular}} & \multicolumn{3}{c}{\textbf{\# Transistors}}     & \multirow{2}{*}{\begin{tabular}[c]{@{}c@{}}\textbf{Power}\\ ($\mu W$)\end{tabular}} & \multirow{2}{*}{\textbf{\# Boxes}} \\ \cline{2-4}          & \multicolumn{1}{c}{\textbf{ON}} & \multicolumn{1}{c}{\textbf{OFF}} & \textbf{Disconnected} &     &           \\ \hline
$3\times 3$        & \multicolumn{1}{c}{1}  & \multicolumn{1}{c}{6}   & 2          & 1.31      & 40000                  \\ \hline
$5\times 5$        & \multicolumn{1}{c}{1}  & \multicolumn{1}{c}{20}  & 4          & 1.48     & 14400                   \\ \hline
$7\times 7$        & \multicolumn{1}{c}{1}  & \multicolumn{1}{c}{42}  & 6          & 1.64      & 7396                   \\ \hline
\end{tabular}
\label{power}
\end{table}

\begin{figure*}[t]
\centering
\includegraphics [width=0.7\linewidth, height= 5cm]{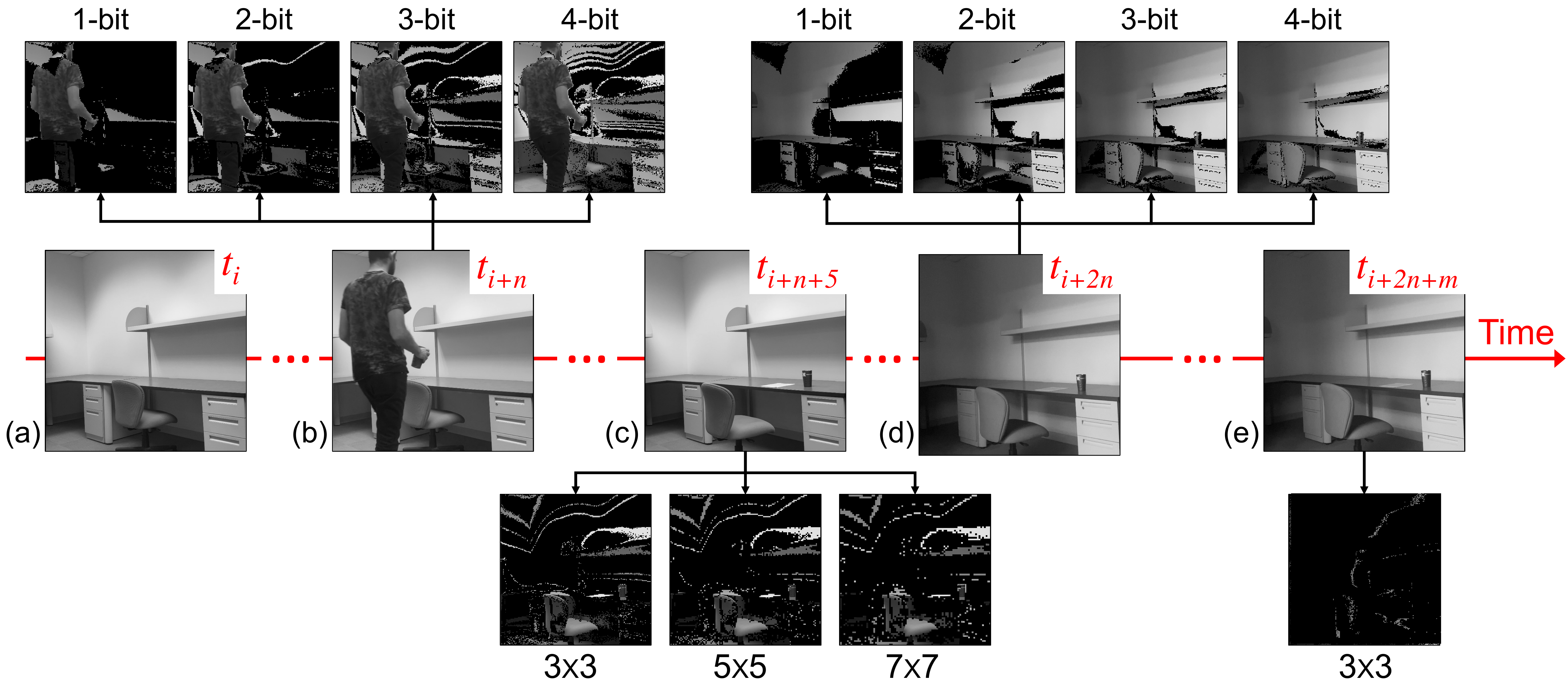} 
\vspace{-0.8em}
\caption{Detecting object timeframes using NeSe, (a) $\xrightarrow{}$(b) detects a person leveraging different precision (1 to 4 bit), (a) $\xrightarrow{}$(c) calculates differences in the images based on different box sizes, (c) $\xrightarrow{}$(d) detects light variation as a new object, and (d) $\xrightarrow{}$(e) updates new background.}
\vspace{-1.5em}
\label{frame}
\end{figure*}

The power consumption and the total number of boxes regarding different box sizes are summarized in Table \ref{power}. Larger box sizes (e.g., $7\times 7$) consist of the lower number of central pixels ($7396$), which leads to more power saving at the cost of accuracy loss.
Another reconfigurable capability of NeSe is the \emph{precision’s} bit-width, which defines the number of compared bits between a pixel and its previous stored value in an MRAM. A lower \emph{precision} requires a smaller number of comparisons and write-back operations that decreases the power consumption but again at the cost of accuracy loss. Thus, a trade-off between efficiency and accuracy can be determined by the user w.r.t available resources, criteria, etc.
Figure \ref{frame} depicts different scenarios, including various box sizes, precisions, light situations, and updating the background. First, NeSe captures Fig. \ref{frame}(a) and stores it as a static background within MRAM cells. Then, an event has occurred in Fig. \ref{frame}(b), and its results related to different precisions are shown. Interestingly, even 1-bit precision removes the background efficiently. Figure \ref{frame}(c) illustrates the results for varied box sizes. After comparing the new input ($t_{i+n+5}$) with the stored background at time $t_i$, we detect that the chair is moved, and the mug is left on the desk. A smaller box size (e.g., $3\times 3$) provides sharper output. After a while, as shown in Fig. \ref{frame}(d) time $t_{i+2n}$, light status changed, but NeSe functions appropriately. Finally, in Fig. \ref{frame}(e), the background is updated by these pixels because the chair locations and the mug remain unchanged for a while. The comparison results using the high accuracy $3\times 3$ boxes exhibit no difference between Fig. \ref{frame}(e) and the new background.
\setlength{\textfloatsep}{0pt}
\begin{algorithm}[t]
\scriptsize
\caption{NeSe Algorithm}
\label{Algorithm1}
\begin{algorithmic}[1]
\State \textbf{Input\textsubscript{2}:} \emph{box\_size} $\in\{3,5,7\}$ \& \emph{precision} $\in\{1,2,3,4\}$-bit 
\State \textbf{Input\textsubscript{3}:} threshold\textsubscript{pixels}, $time_\tau$ 
\State \textbf{Output:} sensor\_mode status
\State      turn\_on\_list = []
\Procedure{Event-Detection}{} 
\State   \textbf{if} time $\geqslant time_\tau$: \Comment{Merge steady objects with the background.}
\State \hspace{0.25cm} \texttt{\textbf{update}} (background)
\State \textbf{for} {$i=\lfloor \frac{\text{\emph{box\_size}}}{2} \rfloor+1~\text{to}~600~\text{with step= \emph{box\_size}}$} 
\State \hspace{0.25cm} \texttt{\textbf{activate}} (row\textsubscript{i})
\State \hspace{0.25cm} pixel\_values $\leftarrow$ \texttt{\textbf{parallel\_read}} (column\textsubscript{i,j}) \Comment{$j\in\{\lfloor \frac{\text{\emph{box\_size}}}{2} \rfloor+1, \dots,600\},~\text{with step= \emph{box\_size}}$}
\State \hspace{0.25cm} num\_changes $\leftarrow$ \texttt{\textbf{parallel\_comp}} (\emph{precision}, pixel\_values, old\_values) 
\State  \hspace{0.25cm} \textbf{if} num\_changes $\geqslant$ threshold\textsubscript{pixels}:
\State \hspace{0.5cm} turn\_on\_list.\texttt{\textbf{push}} ($i$) \Comment{$i$ is row index.}
\State \textbf{if} (length (turn\_on\_list) !=0)
\State \hspace{0.25cm} time += 1 \Comment{Use it to update the background.}
\State	\hspace{0.25cm} \texttt{\textbf{enable}} \textsc{Sensor Mode}
\State \textbf{else}:
\State \hspace{0.25cm} time = 0
\EndProcedure
\Procedure{Sensor Mode}{} 
\While {(length (turn\_on\_list) !=0)}
\State $row$ = turn\_on\_list.\texttt{\textbf{pop}}
\State	\texttt{\textbf{transfer}} ($row - \lfloor box\_size \rfloor \textbf{~to~} row +\lfloor box\_size \rfloor$ )
\EndWhile
\EndProcedure
\end{algorithmic} 
\end{algorithm}

Algorithm\ref{Algorithm1} shows all the steps, including the event-detection and sensing modes provided by the NeSe architecture. The algorithm takes the size of the box, precision, and two thresholds, threshold\textsubscript{pixels}, and time\textsubscript{$\tau$}. The former is used for minimum changes, whereas the latter is leveraged to update the background.
First, every row containing a central pixel, line (9), is activated, and the parallel comparison is performed in line 11 between all the central values and the previous value of the same pixel.
The \texttt{\textbf{parallel\_comp}} function takes the precision, which determines the required number of compared bits. For example, if $precision = 1$, only the most significant bits of pixels are compared. In line 12, if the number of changes is greater than or equal to threshold\textsubscript{pixels}, the row index is held in the turn\_on\_list.
After checking all rows, the length of the turn\_on array is checked. In the case of non-equality to zero, the mode is changed to the sensor mode, and the time counter is increased by one. This variable indicates how many times NeSe is switched to sensor mode continuously. If this variable reaches time\textsubscript{$\tau$}, we need to update the background with the new values (line 7). As shown in Fig.\ref{frame}(e), after updating the background to (d), most of the compared pixels are black.

\subsection{Sensing Mode}
In the sensing mode, all the enabled columns, connected to $V_{dd}$, and rows based on the $R$ signal are connected to ADCs. 
We assume that the background has already been stored in the co-processor, e.g., digital on-chip deep learning accelerator. 
Thus, in the sensing mode, only row indices in the $turn\_on\_list$ should be updated instead of all rows, which results in a considerable power saving.

\section{Performance Evaluation}
\subsection{Power Consumption}
Table \ref{power} reports the power consumption for event detection, i.e., to detect a mismatch between a digitized pixel value and the pre-stored background in MRAM cells assuming two different ADC precisions. The total power consumption for the central pixel comparison can be estimated by 
$P_{total}$= $P_{pixel}$ + $P_{MRAM}$ + $P_{compare}$, where $P_{pixel}$ represents the pixel sensing power that largely depends on the ADC precision. $P_{MRAM}$ is the SOT-MRAM's read power and  $P_{compare}$ denotes the power consumed by the near-sensor CMOS bit-wise XNOR comparator. Assuming a 2-bit ADC structure, every central pixel after readout has to be compared with two SOT-MRAM cells holding the background value. This means 2$\times P_{MRAM}$ are considered in the evaluations. 
We observe that the higher the ADC precision is (here from 4-bit to 2-bit), the higher power budget is required for the edge device to perform such a near-sensor computation and within a particular ADC precision, the larger box size brings higher power efficiency to the system at the cost of lower accuracy as discussed above.

\begin{table}[b]
\caption{Power consumption for event detection w.r.t. ADC precision.}
\centering
\label{power}
\scriptsize
\begin{tabular}{cccc}
\hline
Box size                                                         & 3$\times$3 & 5$\times$5 & 7$\times$7 \\ \hline
\begin{tabular}[c]{@{}c@{}}\# of XNOR (2-bit ADC)\end{tabular} & 80,000     & 28,800     & 14,792     \\
Power (mW)                                                       &  842          & 561.3           &  374.2          \\ \hline
\begin{tabular}[c]{@{}c@{}}\# of XNOR (4-bit ADC)\end{tabular} & 160,000    & 57,600     & 29,584     \\
Power (mW)                                                       &    1,852.4        &  1,234.9          & 823.2           \\ \hline
\end{tabular}
\end{table}

\subsection{Intermittent-Robust Operation}
Power supplies in energy harvesting systems are limited in capacity. Besides, a CMOS-based design loses data when powered down, so restoring (writing back) information after a new power-up consumes power and time. Energy harvesting devices may undergo a charge/discharge cycle hundreds of times per second, which means the system might consume a significant portion of its entire power supply capacity to restore data.
Although NV-MRAMs provide power failure tolerant designs, the required power consumption of write operations for non-volatile elements remains an issue.
Thermal barriers between $40-60~kT$ are generally chosen for MRAM to provide a retention time ($\tau=\tau_0~exp (\Delta/kT)$) of 10-15 years, while the critical spin-current is linearly proportional to the thermal barrier $\Delta$. Thus, for our application that does not require retention times of years, we reduce the thermal barrier of nanomagnets by means of uniaxial anisotropy. 
Herein, MRAM components with $20 kT$ energy barriers are investigated that can achieve retention times ranging from minutes to hours while providing at least 75\% energy reduction.
By reducing the charge currents required for the write operation, significant energy savings can be achieved due to a quadratic relationship between the Ohmic ($I^2R$) losses and the input write currents.


\section{Conclusion}
This paper proposed a practical background subtraction approach, NeSe, for tiny energy-harvested sensors leveraging MRAMs. NeSe allows the accuracy and efficiency of event detection to be adjusted at runtime based on the application's requirements. Furthermore, the proposed design reduces data movement overhead due to the near-sensor implementation of background subtraction. Moreover, MRAMs ensure intermittent resiliency, meaning if the power is cut, the background remains unchanged. Finally, if the moving object is detected, the device switches to the high-powered sensor mode. 

\section*{Acknowledgements}\vspace{-0.5em}
This work is supported in part by the National Science Foundation under Grant No. 2216772 and 2216773.

\bibliographystyle{IEEEtran}

\bibliography{IEEEabrv,./Reference}\vspace{-2em}

\end{document}